\documentclass[reprint,amsmath,amssymb,aps,prl,floatfix]{revtex4-1}
\usepackage{graphicx}
\usepackage{dcolumn}
\usepackage{bm}
\usepackage{multirow}
\usepackage{color}
\usepackage[english]{babel}
\usepackage[T1]{fontenc}
\usepackage{amsmath}
\usepackage{amsfonts}
\usepackage{amssymb}
\usepackage{color}
\usepackage{tabularx,booktabs}
\newcolumntype{Y}{>{\centering\arraybackslash}X}
\usepackage{dcolumn}
\usepackage[unicode=true, bookmarks=true, bookmarksnumbered=false, bookmarksopen=false,
  breaklinks=false,pdfborder={0 0 0},backref=false,colorlinks=true]{hyperref}
 \hypersetup{pdftitle={Half-Heusler alloy LiBaBi: A new topological semimetal with five-fold band degeneracy},
  pdfauthor={Mighfar Imam, Abhishek K. Singh},
  pdfsubject={Computational Material Science},
  unicode=false,pdftoolbar=true,pdfmenubar=true,pdffitwindow=true,pdfstartview={FitH},pdfnewwindow=true,pdfcreator={RevTeX},linkcolor=red,citecolor=blue,urlcolor=blue}
  
\begin{document}

\title{Half-Heusler alloy LiBaBi: A new topological semimetal with five-fold band degeneracy}
\author{Mighfar Imam}
\author{Abhishek K. Singh}
\email{abhishek@mrc.iisc.ernet.in}
\affiliation{Materials Research Center, Indian Institute of Science, Bangalore - 560012, India}
\date{\today}

\begin{abstract}
Based on first-principles study, we report the finding of a new topological semimetal LiBaBi in half-Heusler phase.
The remarkable feature of this nonmagnetic, inversion-symmetry-breaking material is that it consists of only simple $s$- and $p$-block elements. 
Interestingly, the material is ordinary insulator in the absence of spin-orbit coupling (SOC) and becomes nodal-surface 
topological semimetal showing drumhead states when SOC is included. This is in stark contrast to other 
nodal-line and nodal-surface semimetals, where the extended nodal structure is destroyed once SOC is included. Importantly, 
the linear band crossings host three-, four-, five- and six-fold degeneracies near the Fermi level, making this compound very attractive 
for the study of `unconventional' fermions. The band crossing points form a three-dimensional nodal
structure around the zone center at the Fermi level. We identify the surface states responsible for the appearance of the drumhead states.
The alloy also shows a phase transition from topological semimetal to a trivial insulator on application of pressure.
In addition to revealing an intriguing effect of SOC on the nodal structure, our findings introduce 
a new half-Heusler alloy in the family of topological semimetals, thus creating more avenues for experimental exploration.
\end{abstract}

\maketitle

\section{Introduction}

In recent years, topological materials have become the focus of intense research in condensed matter physics 
and materials science, since they exhibit fundamentally new physical phenomena with potential applications for novel devices \cite{Bernevig1757, MooreNature464.2010, 
PhysRevB.83.205101,PhysRevB.84.075129,PhysRevLett.107.186806, PhysRevB.87.245112}. 
The first three-dimensional topological materials to be predicted and subsequently discovered were topological 
insulators \cite{MooreNature464.2010, hasan2010colloquium,RevModPhys.83.1057, PhysRevLett.98.106803, Hsieh919}. Topological insulators have conducting 
surface states while they are insulating in bulk, and they are characterized by the so-called $Z_2$ topological invariant associated with the 
bulk electronic structure. Their band structures are usually characterized by a band inversion that involves the switching of bands of opposite parity
around the Fermi level. After the groundbreaking discovery of topological insulators, however, the 
experimental observations of Weyl semimetals \cite{PhysRevB.83.205101, BurBal11, PhysRevB.86.115208, 
weng2015weyl, xu2015discovery, Xu294, PhysRevLett.107.186806,HuaXuBel15} has partially shifted the research interest 
from insulating materials to semimetals and metals.

The topological semimetals have been characterized based on dimensionality and degeneracy of band crossings: 
a zero-dimensional crossing (nodes) with two- and four-fold band degeneracy defines the Weyl and Dirac semimetals \cite{PhysRevLett.108.140405, wang2012dirac, PhysRevB.88.125427} respectively, 
while a one-dimensional crossing gives the corresponding nodal-line semimetals \cite{PhysRevLett.115.036806,yu2015topological}. In all of these, the band crossing points are 
formed due to band inversion \cite{YanNag14}. The band crossing points
of Weyl semimetals have definite chirality and they are located 
at an even number of descrete points in the Brillouin zone (BZ). The characteristic feature of the Weyl semimetals is the existence of special surface states, known as fermi arcs, whose end points terminate
on (the surface projection of) a pair of nodes with opposite chirality. Dirac semimetals with four-fold band degeneracy can be thought of as three-dimensional analogue of 
graphene. Nodal-line semimetals are generally considered as precursor states for other topological states: they might evolve into Weyl points, convert into Dirac points, or become a 
topological insulator by introducing the spin-orbit coupling (SOC) or mass term \cite{Yu2016}. Very, recently, however, `new fermions' beyond the above mentioned conventional 
ones (which are characterized by two- or four-fold crossings), have been identified, which have expanded the classification of fermions in crystal lattice  \cite{PhysRevLett.116.186402,Bradlynaaf5037,
PhysRevLett.117.076403,PhysRevB.93.241202,PhysRevX.6.031003,PhysRevB.94.165201,2016arXiv160506831C}. While many studies have predicted the existence of a three-fold 
degeneracy-- which has also been confirmed experimentally \cite{2016arXiv161008877L}-- the existence of unconventional fermions with three-, six- and eight-fold band 
degeneracies that appear at high-symmetry points in non-symmorphic crystals has been predicted in Ref \cite{Bradlynaaf5037}.

In this study we identify a half-Heusler compound LiBaBi, which shows a hitherto unknown
fermion with five-fold band degeneracy-- in addition to three-, four- and six-fold -- on the two-dimensional surface of the Brillouin zone.
The band crossing accompanied by the band inversion, results from the inclusion of SOC. 
We show the presence of special surface states giving rise to drumhead states, thus confirming the
existence of nontrivial topology in this material. We also study the effect of pressure on the band topology and show that the application of pressure of about 1 GPa
causes the phase transition from the topological semimetal phase to trivial insulating phase.

\section{Methodology}

The calculations were performed using density functional theory (DFT) \cite{kohn1965self} as 
implemented in the Vienna ab initio simulation package (VASP) \cite{kresse1996efficiency,
kresse1996efficient}. Projector augmented wave (PAW) \cite{blochl1994projector,kresse1999ultrasoft} 
pseudopotentials were used to represent the ion-electron interactions. The exchange and 
correlation part of the total energy was approximated by the generalized gradient approximation
 (GGA) using Perdew-Burke-Ernzerhof (PBE) type of functionals \cite{perdew1996generalized}. The 
wave functions were expanded in a plane wave basis with a high energy cut-off of 400 eV and a 
Monkhorst-Pack \cite{monkhorst1976special} \textbf{k}-grid of 12$\times$12$\times$12 in the 
Brillouin zone. The calculations were done both with and without spin-orbit coupling. 
The surface spectrum including the surface bands and Fermi surfaces 
were calculated based on the iterative Green's function method \cite{sancho1985highly} 
after obtaining the tight-binding Hamiltonian from the maximally localized Wannier functions 
\cite{mostofi2014updated}, as implemented in the WannierTools package\cite{wann_tools}.

\section{Results and Discussion}
\subsection{}

\begin{figure*}[ht!]
\includegraphics[width=0.9\textwidth]{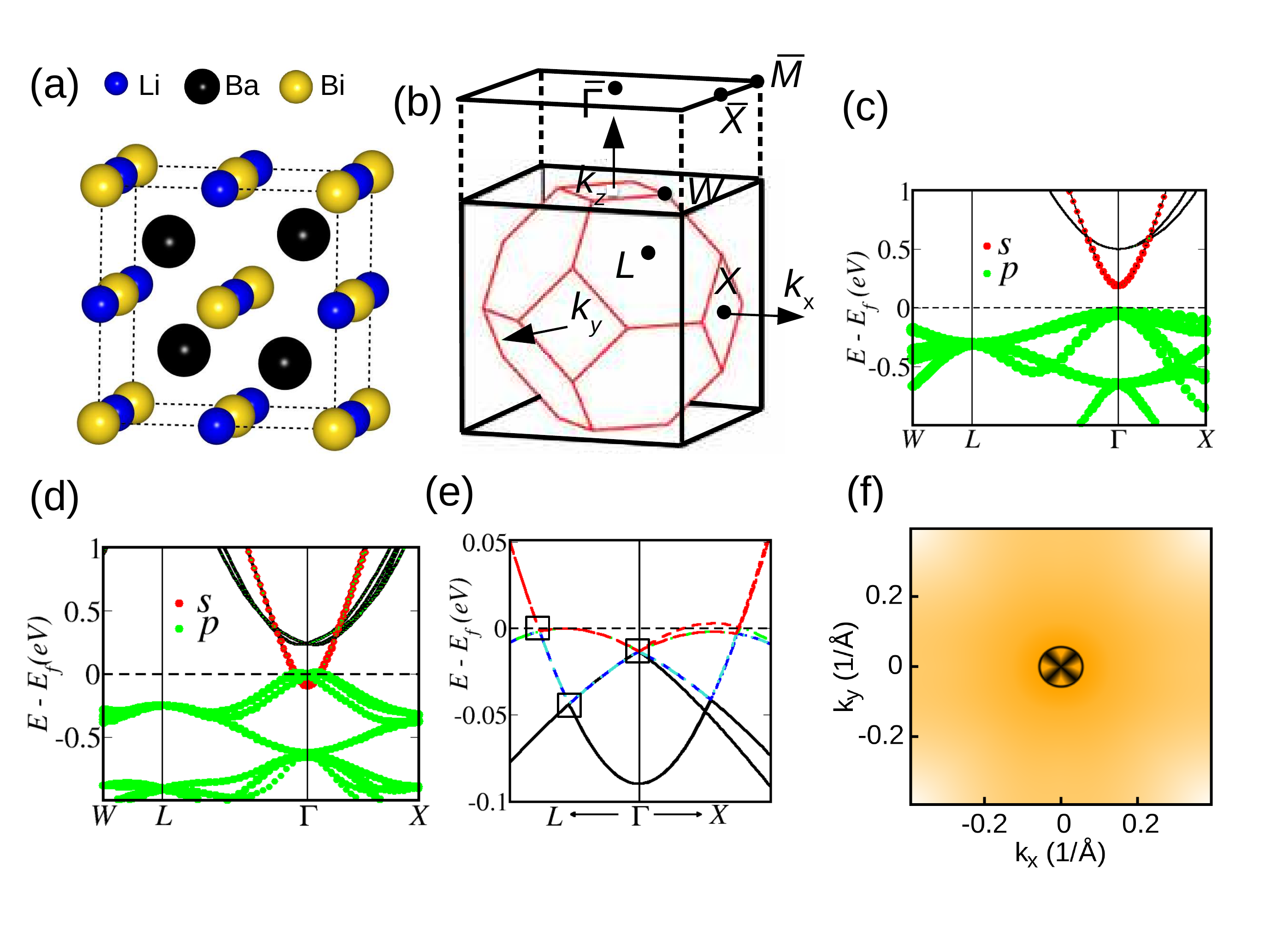}
\vspace {-1cm}
\caption{(Color online) (a) The crystal structure of LiBaBi compound in half-Heusler phase. (b) The 
three-dimensional Brillouin zone of the structure and the projected plane showing the  
two-dimensional surface Brillouin zone to the (001) surface. The band structure plots along the high symmetry
directions (c) without spin-orbit coupling, and (d) with spin-orbit coupling. The orbital
contributions are represented by red and green filled circles for the $s$- and $p$-orbitals 
respectively. (e) The enlarged view of the band touching region in (d).  (f) Nodal-surface
structure shown in the $k_z$=0 plane. The dark color represents the regions where the lowest 
conduction band and the highest valence band touch each other.}
\label{fig:1}
\end{figure*}

The Heusler family trademark is a simple structural framework that can accommodate a vast number of
different element combinations, resulting in phases with a large diversity in physical properties. In particular, 
this big family of tunable, multifunctional compounds may substantially enlarge the number of topological materials known.
Half-Heusler materials have so far mostly been identified as potential topological insulators 
\cite{LinWraXia10, ChaQiKub10} but not much as topological semimetals. Very recently, however, 
Weyl semimetals have been reported in Co-based magnetic full- \cite{WanVerKus16} 
and half-Heusler alloy GdPtBi \cite{HirKusWan16}. These Heusler materials still require the 
presence of some $d$-block or heavy metal in order to show topological features.
In the half-Heusler phase XYZ, X is typically a heavy transition metal, Y is a light transition metal
or a rare-earth metal, while Z is a late $p$-block (main-group) element. However, in the present study, in order to 
explore new possibilities, we have changed the usual picks for X and Y and considered instead simple metals from $s$-block. He we present our study on 
LiBaBi with Li, Ba and Bi taken from groups I(A), II(A) and V(A) respectively. 

Fig. \ref{fig:1}(a) shows the crystal structure of half-Heusler alloy LiBaBi in symmorphic space group $F\bar{4}3m$ (No. 216). Its unit cell contains four formula units. 
Li, Ba and Bi atoms occupy the following three crystal sites of the 
cubic lattice: (1/2,1/2,1/2), (1/4,1/4,1/4) and (0,0,0), respectively in Wyckoff 
coordinates. The site of Ba atoms breaks the inversion symmetry. The optimized lattice constant of LiBaBi is found to be 7.99 \AA. The internal atomic 
coordinates were fixed at the symmetry sites.

Fig. \ref{fig:1}(b) shows the bulk Brillouin zone and the projection of the (001) surface Brillouin zone. Fig. \ref{fig:1}(c) shows the 
band structure of LiBaBi along the high symmetry  lines in the BZ without spin-orbit coupling. The contribution
from the $s$- and $p$-like states in the band structure is shown with red and green balls respectively. 
Without SOC, the system is a trivial insulator with normal band ordering of $s$- and $p$-character 
states forming the conduction band minimum (CBM) and valence band maximum (VBM), respectively. At 
$\Gamma$-point, the VBM is formed exclusively by $p$-states of Bi while the CBM comes predominantly from $s$-states 
of Bi and Li. The lowest conduction band is two-fold degenerate while the highest valence bands, 
at the $\Gamma$-point, are six-fold degenerate. There is a direct bandgap of about 0.23 eV at the $\Gamma$-point.

Turning on SOC brings about important changes in the band structure near the Fermi energy (Fig. \ref{fig:1}(d)).
Around the $\Gamma$-point, the bottom of the conduction band moves down and crosses the 
Fermi level to form  electron pocket and similarly the top of the valence bands moves up 
to form tiny hole pockets. This results in a band inversion at $\Gamma$-point with the $s$-character 
states coming below the $p$-character states. The band inversion is one of the key ingredients for
topological materials \cite{RevModPhys.83.1057}. A blow-up of the band crossing regions (Fig. \ref{fig:1}(e)) 
shows that some bands actually do not cross but instead form avoided crossings, as a result of SOC. We call these avoided crossing points as 
band touching points (BTP), some of them marked with squares in Fig. \ref{fig:1}(e). The avoided 
crossing width (SOC splitting) is very small (about 1--2 meV) and therefore the BTPs are considered degenerate. At some BTPs, 
one band actually crosses from lower band to upper band. The band dispersions around the BTPs are linear. There is a BTP with six-fold 
degeneracy at the $\Gamma$-point, slightly below the Fermi level. Along the $\Gamma$-L 
direction, the conduction and valence bands stick together before they reach another BTP (located at the Fermi energy)
with five-fold degeneracy, shown by the left-most point marked with square in Fig. \ref{fig:1}(e). There is another five-fold 
degenerate BTP below this point, which is closer to the $\Gamma$-point.
These BTPs with five-fold degeneracy have one band which crosses from lower to upper band, making these points truly gapless. Similarly along the $\Gamma$-X direction, there are BTPs but with three- or four-fold degeneracies. The reduced
degeneracy along the $\Gamma$-X direction is due to splitting between one of the three $p_x$, $p_y$ and $p_z$ bands.
This particular structure of band crossings results in a finite Fermi surface (at any chemical potential) which can not 
be reduced to a point by breaking some crystalline symmetry. This is similar to $\theta$-TaN phase topological semimetal \cite{PhysRevB.93.241202}.
The degeneracy of the band crossing along the $\Gamma$-L is protected by three-fold rotational symmetry about the body diagonal.

This $\Gamma$-centered structure of band touchings is along all the (equivalent) directions in the 3D BZ, forming a spherical nodal-surface.
A two-dimensional slice of the spherical nodal surface is shown in Fig. \ref{fig:1}(f) for $k_z$=0 plane. The dark color represents
the regions where conduction and valence bands touch each other. Centered around the $\Gamma$-point is the circular ring-shaped structure,  
which forms the spherical nodal surface in the 3D BZ. This nodal surface hosts different degeneracies in different directions, 
as discussed in the band structure.

Topological nodal-line and nodal-surface semimetals form a distinct class of topological materials 
beyond topological insulators, Weyl or Dirac semimetals. The effect of SOC in general has been detrimental to the stability of nodal-surface and nodal-lines 
\cite{PhysRevLett.115.036806,PhysRevLett.116.195501,BzdWuRue16, PhysRevB.92.045108,PhysRevLett.115.036807,PhysRevB.95.014418,PhysRevB.95.045136}. However, 
in some materials, nodal-lines are obtained in the presence SOC \cite{PhysRevB.93.121113,
PhysRevB.93.085427, BiaChaSan16}. This is, to the best of our knowledge, the first system where SOC 
brings about a stable nodal surface which was absent when SOC was not included. Also, the topological semimetals with extended (non-point) crossings so far have been discovered for either two-fold or four-fold degenerate bands
giving rise to nodal-line or Dirac nodal-line semimetals, respectively. Our nodal surface hosts new degeneracy of three and five.
In Ref \cite{Bradlynaaf5037}, the authors have studied the topological features of fermions arising from 
three-, six- and eight-fold band crossings. However, the nature of band degeneracies in our system
is different, namely, their occurrence at non-high-symmetry points and lack of nonsymmorphic symmetry. Moreover, the 
compound LiBaBi with five-fold band degeneracy in a symmorphic space group is new and not studied so far.

\begin{figure*}
\includegraphics[width=0.9\textwidth]{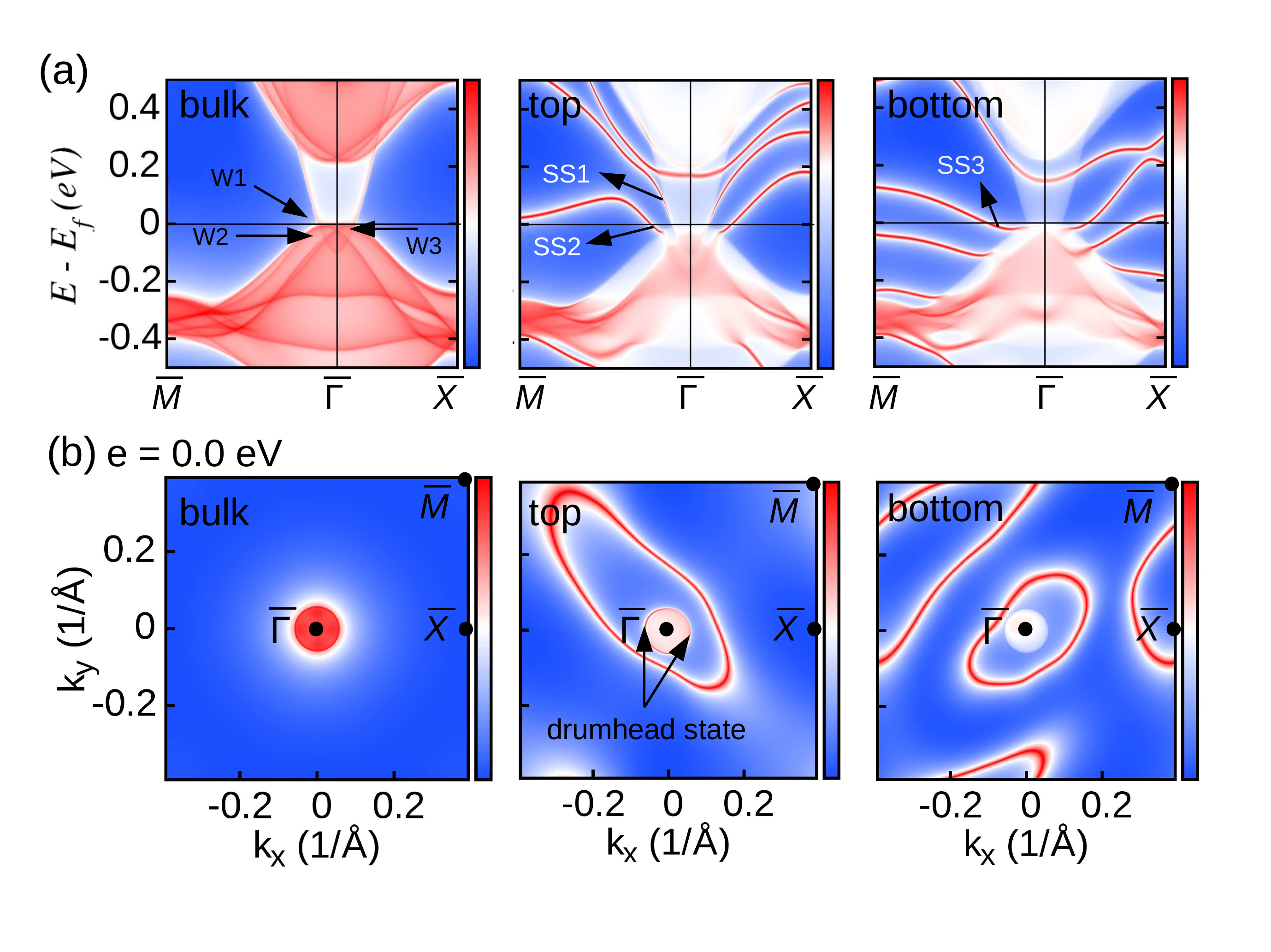}
\vspace {-1cm}
\caption{(Color online) Projected surface band dispersions and their Fermi surfaces. (a): band dispersions along high 
symmetry directions of the surface Brillouin zone for the projected bulk states (left panel), 
top and bottom surfaces (middle and right panels respectively).
(b): Constant-energy cut through the entire surface Brillouin zone for the energy e= 0.0 eV 
(Fermi level). Projected bulk Fermi surface (left panel) and surface states on the top (middle panel)
 and bottom surfaces (right panel). The spectral density is represented as color 
scale: blue: minimum, red: maximum (i.e., the red color for the left panels in (a) and (b) represents
bulk states, whereas for the middle and right panels, represents surface states). High-symmetry
points on the surface Brillouin zone are marked with black dots in (b).}
\label{fig:2}
\end{figure*}

The band inversion with linear band crossings near the Fermi energy together with nodal-surface suggest the presence of topologically non-trivial
surface states in this compound. Nodal-line semimetals produce the so-called drumhead-like surface states either nestled inside or dispersing outside of 
the projected nodal-loop \cite{burkov2011topological,PhysRevB.92.045108,yu2015topological,BiaChaSan16,PhysRevB.93.121113,PhysRevB.93.201114,PhysRevB.95.014418}.
Fig. \ref{fig:2} shows the band dispersion and surface state electronic structure (spectral intensity maps) 
projected on the BZ surfaces perpendicular to the $k_z$ axis. Since a slab calculation involves two surfaces, 
the corresponding surface bands and spectral intensity maps for both surfaces are given. In the middle and right panels of Fig. \ref{fig:2}(a), the sharp red curves 
represent the surface states whereas the shaded regions show the spectral weight of projected bulk states. Unlike the projected bulk band which is independent of the presence of surface, 
the surface bands are sensitive to the surface orientation (top or bottom). However, for both the surfaces, there are surface states which appear in the bulk 
band gap region while originating from the bulk BTPs (indicated with arrows in the left panel). This indicates the topological nature of the bulk nodal structure. The linearly dispersing 
surface state SS1, appearing on the top surface (middle panel), originates from the nodal point W1, while the surface states SS2 (middle panel) and SS3 (right panel) 
seem to originate from points W2 and W3, respectively. The latter two states are masked by the bulk Fermi surface around the $\Gamma$-point, hence their originating points are not visible. The surface state 
SS1 disperses outwards with respect to $\Gamma$-point and merges with bulk bands near the node. Unlike the nearly flat drumhead-like states
in some previous studies \cite{PhysRevB.92.045108,yu2015topological}, these states are rather dispersive, as also found in some recent works \cite{PhysRevB.95.045136,PhysRevB.93.201114, PhysRevB.93.121113}.
The difference in surface electronic structure for the top and bottom surfaces indicates the absence of inversion symmetry in the system.

Fig. \ref{fig:2}(b) shows the corresponding Fermi surface at the Fermi energy.
The left panel in Fig. \ref{fig:2}(b) shows the projected bulk Fermi surface which is given by the circular region bounded by the BTPs W1 around the zone center. 
The projected drumhead surface state around the circular loop (the projection of the spherical nodal-surface onto the surface BZ) is clearly seen as red circle in the middle panel in Fig. \ref{fig:2}(b).
The linearly dispersing state SS1 produces this circular state on the top surface while there is no such state on the bottom surface (right panel). 
The Fermi state forming the big closed-loop around the $\Gamma$-point on the top surface comes 
from the state SS2 while the one on the bottom surface comes from the state SS3. These loops are elongated in the $k_x =\pm k_y$ directions.

\begin{figure}
\includegraphics[width=\columnwidth, trim={0 0 6cm 0},clip]{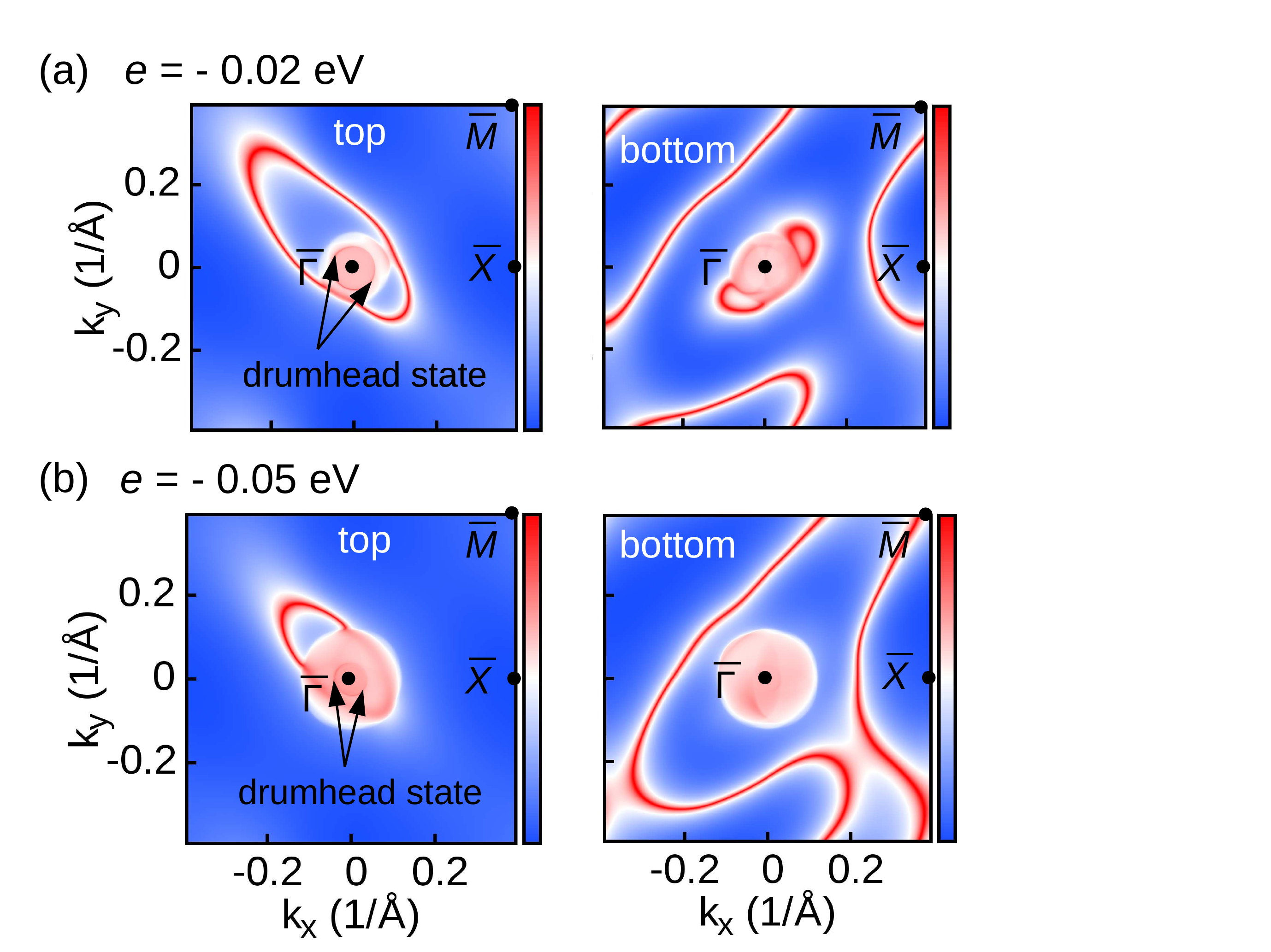}
\caption{(Color online) Constant-energy cuts through the entire surface Brillouin zone for the energies (a) e = -0.02 eV
and (b) -0.05 eV corresponding to the bulk band touching points W2 and W3 in Fig. \ref{fig:2} (a:left panel). The left and right panels represent the top and bottom surfaces respectively. High-symmetry
points on the surface Brillouin zone are marked with black dots.}
\label{fig:3}
\end{figure}

In order to see the topology of the Fermi surface at other band crossings, we plot in Fig. \ref{fig:3} the surface states at the chemical potentials corresponding 
to BTPs W2 and W3. As one goes down in binding energy, the projected bulk Fermi surface
becomes bigger due to the peculiar band dispersion around the zone center. As can be seen in Fig. \ref{fig:3}, the circular Fermi state on the projected nodal line (the top surface) is present at 
lower binding energies as well. Note that at -0.05 eV energy, corresponding to the five-fold BTPs W2, the circular Fermi state surrounds a smaller circle (Fig. \ref{fig:3}(b) left panel), which is the projection 
of the nodal surface formed by BTPs W2. The surface states SS2 and SS3 give rise to Fermi loops at these energies also. The disjoint Fermi arc-type states in Fig. \ref{fig:3}(a) 
right panel and (b) left panel, produced by SS3 and SS2 respectively, are actually closed loops but the other part of the loops are masked by the bulk Fermi surface.

Presence of drumhead states in LiBaBi suggests that the topology of five-fold crossings are similar to that of two-fold crossings of (Weyl) nodal-line semimetals rather than of Dirac nodal-line semimetals.

\begin{figure}
\includegraphics[width=0.5\textwidth]{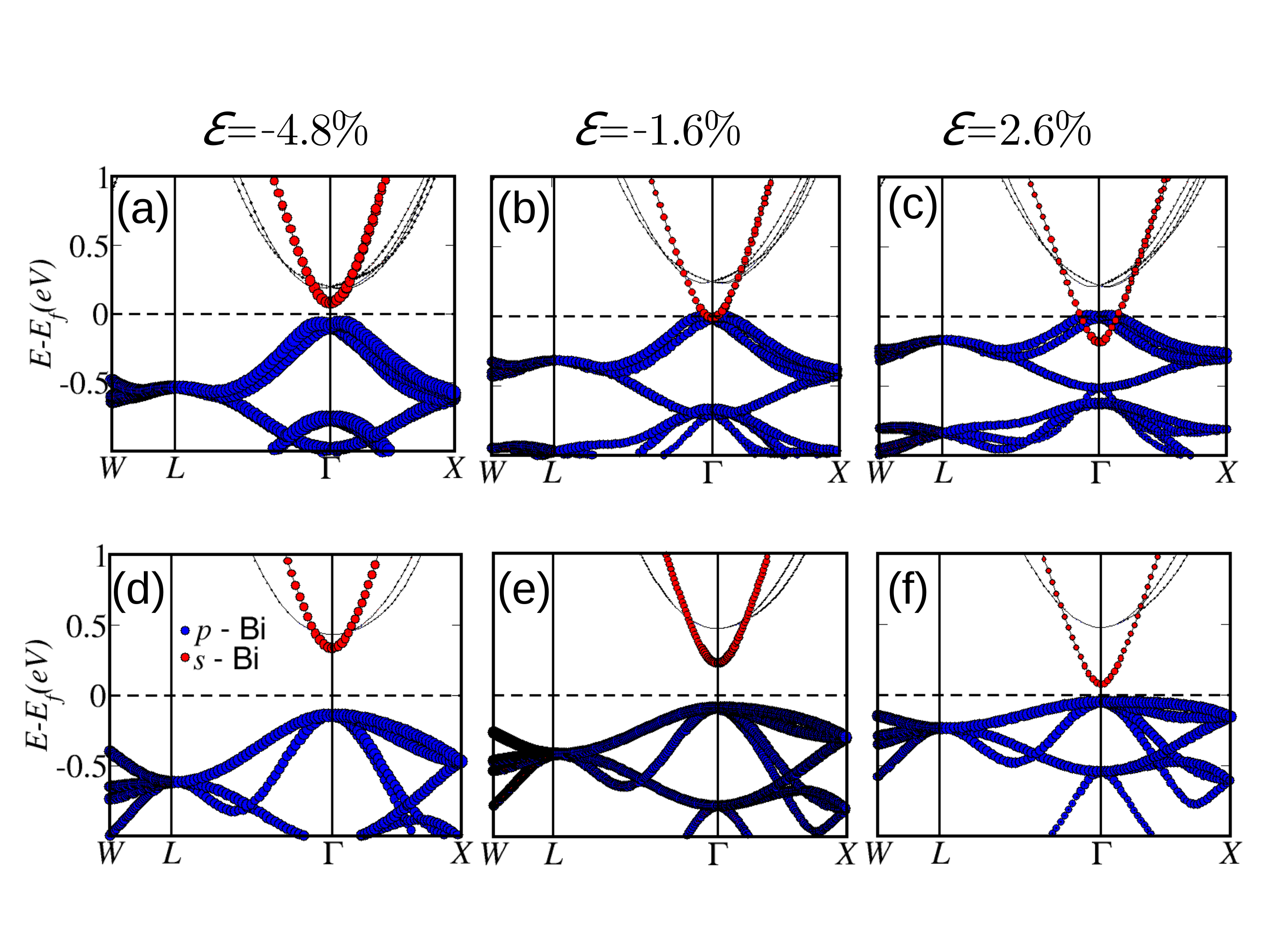}
\vspace {-1cm}
\caption{(Color online) Band structure plots at different lattice strains ($\varepsilon$) with (upper panel) and without (lower panel) SOC included. The blue and red balls represent the $p$- and $s$-character 
states of Bi respectively. The transition point is at $\varepsilon$ = -1.6\%, where the conduction and valence bands begin to touch each other near the $\Gamma$-point (with SOC). The strain values of -4.8\%, -1.6\%
and 2.6\% correspond respectively to pressure values of 1.46 GPa, 0.37 GPa and -0.67 GPa.}
\label{fig:4}
\end{figure}

This material also shows a pressure-induced phase transition. Fig. \ref{fig:4} shows the band structure plots at different lattice strains. At a compression of 
about 4.8\% (Fig. \ref{fig:4}(a)), which is equal to a hydrostatic pressure of about 1.46 GPa, the VBM and CBM move away from each other, reversing the band order 
and creating a band gap at the $\Gamma$-point. This makes the compound a trivial 
insulator with calculated $Z_2$ invariant equal to zero. With respect to the equilibrium lattice constant where the valence and conduction bands have already crossed each other 
(Fig. \ref{fig:1}(d)), the transition point, where these bands just begin to touch, is situated at a smaller lattice constant (Fig. \ref{fig:4}(b)). However, when the lattice is stretched 
(Fig. \ref{fig:4}(c)), the system remains a topological semimetal with the inverted conduction band at the $\Gamma$-point going further down in energy. Hence, this compound would show a phase transition from a
topological nodal-surface semimetal to a trivial insulator with the application of hydrostatic pressure. At the transition point, LiBaBi is expected to show the emergence of nontrivial surface states. 
The corresponding bands without SOC (Fig. \ref{fig:4}(d)-(e)) show trivial insulator phase. As the pressure is reduced, the band gap decreases. At larger lattice stretching (not shown here), 
the non-SOC bands would also show a band inversion and transition to topological phase. In this way, the inclusion of SOC advances the onset of topological phase transition and makes it to occur at the 
equilibrium lattice constant (at zero pressure). The inclusion of SOC is, therefore, crucial to model real topological materials.



\section{Conclusion}
In conclusion, we have shown a new topological nodal-surface semimetal with five-fold band degeneracy in a half-Heusler compound LiBaBi
consisting only of $s$- and $p$-block elements. The formation of the nodal-surface is brought about by the inclusion of spin-orbit coupling, without which, the material is a trivial insulator.
Similar to nodal line semimetals, this material shows drumhead-type Fermi states around the projected nodal-line. A new feature of the nodal structure is that the 
band crossings are five-fold degenerate, making this material having important consequences for `unconventional fermions'. Further, we show the effect of straining the lattice
on the band topology of the material. A compressive pressure causes the compound to become a trivial insulator while the material remains a topological semimetal for expansive pressures.
Our findings thus introduce a new material and extend the domain of search for practical topological semimetals.

\section{Aknowledgement}
The authors thank Dr. Swaminathan Venkataraman
for valuable discussions. This work is partly supported
by the U.S. Army Contract FA5209-16-P-0090. We also
acknowledge Materials Research Center and Supercomputer Education and Research Centre, Indian Institute
of Science for providing computational facilities.

\bibliography{references}

\end{document}